\documentclass[reprint, aps, superscriptaddress, citeautoscript, floatfix, amsmath, amssymb, prb]{revtex4-1}

\usepackage{lmodern,microtype}
\usepackage[mathlines]{lineno}
\relax

\usepackage{xcolor}
\usepackage{lipsum}
\usepackage{graphicx}
\usepackage[caption=false]{subfig}
\usepackage{array,booktabs,paralist,dcolumn}
\usepackage{enumitem}
\usepackage{multirow}
\usepackage{mathtools,amssymb}
\usepackage{mathdesign,bm}

\usepackage{hyperref}
\usepackage{filecontents}

\setlist[itemize]{leftmargin=*}

\begin{document}
\title{Phonon, Electron, and Magnon Excitations in Antiferromagnetic L1$_{0}$-type MnPt}

\author{Kisung Kang}
\affiliation{Department of Materials Science and Engineering, University of Illinois at Urbana-Champaign, Urbana, IL 61801, USA}

\author{David G. Cahill}
\affiliation{Department of Materials Science and Engineering, University of Illinois at Urbana-Champaign, Urbana, IL 61801, USA}
\affiliation{Materials Research Laboratory, University of Illinois at Urbana-Champaign, Urbana, IL 61801, USA}

\author{Andr\'{e} Schleife}
\email{schleife@illinois.edu}
\affiliation{Department of Materials Science and Engineering, University of Illinois at Urbana-Champaign, Urbana, IL 61801, USA}
\affiliation{Materials Research Laboratory, University of Illinois at Urbana-Champaign, Urbana, IL 61801, USA}
\affiliation{National Center for Supercomputing Applications, University of Illinois at Urbana-Champaign, Urbana, IL 61801, USA}

\begin{abstract}
Antiferromagnetic L1$_{0}$-type MnPt is a material with relatively simple crystal and magnetic structure, recently attracting interest due to its high N{\'{e}}el temperature and wide usage as a pinning layer in magnetic devices.
While it is experimentally well characterized, the theoretical understanding is much less developed, in part due to the challenging accuracy requirements dictated by the small underlying energy scales that govern magnetic ordering in antiferromagnetic metals.
In this work, we use density functional theory, the Korringa-Kohn-Rostoker formalism, and a Heisenberg model to establish a comprehensive theoretical description of antiferromagnetic L1$_{0}$-type MnPt, along with accuracy limits, by thoroughly comparing to available literature data.
Our simulations show that the contribution of the magnetic dipole interaction to the magnetocrystalline anisotropy energy of $K_{1}$=1.07$\times 10^{6}$\,J/m$^3$ is  comparable in magnitude to the spin-orbit contribution.
Using our result for the magnetic susceptibility of $5.25\times10^{-4}$, a lowest magnon frequency of about 2.02\,THz is predicted, confirming THz spin dynamics in this material.
From our data for electron, phonon, and magnon dispersion we compute the individual contributions to the total heat capacity and show that the dominant term at or above 2\,K arises from phonons.
From the Landau-Lifshitz-Gilbert equation, we compute a N\'{e}el temperature of 990--1070 K.
Finally, we quantify the magnitude of the magneto-optical Kerr effect generated by applying an external magnetic field.
Our results provide insight into the underlying physics, which is critical for a deep understanding of fundamental limits of the time scale of spin dynamics, stability of the magnetic ordering, and the possibility of magneto-optical detection of collective spin motion.
\end{abstract}

\maketitle

\section{\label{sec:intro}Introduction}
Several decades after their initial discovery \cite{Neel:1948}, antiferromagnetic materials are recently attracting great interest, owing to the successful probing and manipulation of their magnetic ordering by electrical and optical means.
Electrical switching of antiferromagnetic CuMnAs was reported \cite{Wadley:2016} in 2016 and the switching of the N\'{e}el vector was concluded from measuring the magneto-optical Voigt effect \cite{Saidl:2017}.
Electrical read-out was demonstrated for antiferromagnetic Mn$_2$Au using anisotropic magnetoresistance \cite{Meinert:2018}.
In addition, while ferro- or ferri-magnets are easily affected by external fields, collinear antiferromagnets are robust against such manipulation due to their vanishing net magnetization.
This initially hampered applications, however, it has now become the reason for the use of antiferromagnets as excellent pinning layers:
They maintain their magnetic ordering under external fields, while providing strong exchange bias on the adjacent ferromagnetic or ferrimagnetic layers \cite{Nojuas:1999}.

The material investigated in this work, antiferromagnetic L1$_{0}$ type MnPt, follows a similar timeline:
Based on neutron powder diffraction Andersen \emph{et al.}\ explained \cite{Andresen:1965} its magnetic structure as early as 1965 invoking antiferromagnetically and ferromagnetically coupled moments along the [110] and [001] directions, respectively, and a N\'{e}el vector orientation along [001].
The potential for spin-flip transitions of the magnetic alignment from [001] to [100] was recognized from two different neutron scattering experiments on powder samples \cite{Kren:1968, Severin:1979}.
In addition, also single crystal neutron scattering measurements recently confirmed this spin-flip transition between 580\,K and 770\,K, aligning the moments along [100] \cite{Hama:2007}.
Finally, a relatively high N\'{e}el temperature of 970\,--\,975\,K was measured for MnPt \cite{Kren:1968,Umetsu:2006}, causing its magnetic properties to be thermally stable at room temperature.
Applications of antiferromagnetic MnPt include spin-valve structures with giant magnetoresistance, based on exchange bias at the interface with a ferromagnetic layer \cite{Anderson:2000, Toney:2002, Rickart:2004}, and there is increased interest in this material as pinning layer in devices \cite{Anderson:2000}.
However, the fundamental exchange interactions are still under the veil, preventing detailed theoretical understanding of the N\'{e}el temperature or the wave vector dependent magnon dispersion, which also contributes to the heat capacity.

While the experimental characterization of structural and magnetic properties of MnPt is thorough, the theoretical understanding is much less developed.
Metallic AFMs constitute a challenge in particular for first-principles simulations since the underlying energy scales oftentimes push the accuracy of numerical convergence to its limits.
The relatively simple chemical structure and magnetic configuration make MnPt an ideal candidate to explore this issue for first-principles simulations of ground- and excited-state properties.
In this work, we establish a thorough comparison between our first-principles data, other computational data from the literature, and experiments, to discuss reasons for deviations.

First, we study the atomic geometry of MnPt, its magnetic structure, and susceptibility.
We then compute exchange parameters to model the magnetic structure, confirming the early analysis by Andersen \emph{et al.}\ \cite{Andresen:1965}
Our simulations of the magnon gap and the N\'{e}el temperature are in good agreement with experimental data\cite{Kren:1968, Umetsu:2006}.
For magnetocrystalline anisotropy, which helps to explain the orientation of the N\'{e}el vector in the ground state and to understand barriers against its reorientation, we compare our data with prior first-principles results and identify a so far overlooked classical contribution due to magnetic dipole interactions.
Our analysis forms a basis of future studies, e.g.\ of the strain dependence of magnetic ordering and magnetocrystalline anisotropy, possibly helping to explain reports of non-volatile modulation of resistance using piezoelectric strain, with possible application in strain-induced switching \cite{Yan:2019}.

Furthermore, our simulations provide predictions that enable deeper understanding of the underlying physics of antiferromagnetic L1$_{0}$-type MnPt:
This includes fundamental limits to the time scale of spin dynamics, the thermal stability of the antiferromagnetic ordering at room temperature,
the relative contributions of electrons, phonons, and magnons to the heat capacity of this material, and the potential for using MnPt for magneto-optical detection of collective spin motion via MOKE measurements of precession.
To this end, we derive spin dynamics and N\'{e}el temperature from the Landau-Lifshitz-Gilbert equation.
We predict excited-state properties such as phonon, electron, and magnon dispersion:
The electronic band structure is approximated by Kohn-Sham eigenvalues, the phonon dispersion is computed within the frozen phonon approximation, and the magnon dispersion is obtained from linear spin-wave theory.
We use this data to compute the total heat capacity of the material, in good agreement with experiment\cite{Umetsu:2006}, and also directly compare the electronic heat capacity to data from thermal relaxation experiments \cite{Umetsu:2006}.
Our results show that up to about 2 K, there are electronic contributions to the total heat capacity, but at higher temperatures most of the total heat capacity originates from phonons instead of magnons, due to the magnon gap and the low magnon density of states.
This is different from materials with magnetic critical temperatures of just a few K \cite{Khuntia:2010}, for which the magnon heat capacity can be larger than the phonon heat capacity at low temperatures.
Comparing the individual contributions to the heat capacity, computed from the energy dispersion relations of phonons, electrons, and magnons, to experiment provides insight into the relative accuracy of our first-principles results.
Finally, from the electronic band structure, including spin-orbit coupling, we predict optical and magneto-optical spectra of antiferromagnetic MnPt, explaining the relative importance of contributions from Mn and Pt.

After introducing the computational approaches in Sec.\,\ref{sec:comp}, the ground state properties of antiferromagnetic L1$_{0}$-type MnPt are discussed in Sec.\,\ref{sec:grnd}, including relaxed atomic coordinates and magnetic structure, magnetocrystalline anisotropy energy, magnetic susceptibility, and exchange coupling parameters.
In Sec.\,\ref{sec:enrg}, we report first principles results for the dispersion relations of electrons, phonons, and magnons, and discuss their individual contributions to the total heat capacity, which we also compare to experiment.
Finally, in Sec.\,\ref{sec:exct} we report the N{\'{e}}el temperature and analyze optical and magneto-optical properties in detail.

\section{\label{sec:comp}Computational Details}
First-principles simulations of MnPt are carried out within density functional theory (DFT), as implemented in the Vienna \emph{Ab-Initio} Simulation Package (\texttt{VASP}) \cite{Kresse:1996,Kresse:1999,Gajdos:2006,Steiner:2016}.
Exchange and correlation is described by the generalized-gradient approximation developed by Perdew, Burke, and Ernzerhof (PBE) \cite{Perdew:1997}.
Kohn-Sham states are expanded into plane waves up to a kinetic-energy cutoff of 600 eV.
A $15\,\times15\,\times15$ Monkhorst-Pack (MP) \cite{Monkhorst:1976} $\mathbf{k}$-point grid is used to sample the Brillouin zone for structural relaxation and optical spectrum calculations, leading to total energies that are converged to within $0.1$ meV/atom.
Computing the anisotropy energy requires a denser $24\,\times24\,\times24$ MP $\mathbf{k}$-point sampling to converge the anisotropy energy within $0.03$ meV/atom.
Each self-consistent calculation is performed for collinear (atomic relaxations) or noncollinear (optical properties with tilted magnetic moments) magnetic ordering first, neglecting the spin-orbit interaction.
Subsequently, spin-orbit coupling is described non-selfconsistently, by using the resulting Kohn-Sham states and charge density to set up the Kohn-Sham Hamiltonian and diagonalizing it including the spin-orbit coupling term.
From this, we compute ground state energies and optical properties.

We further compute phonon frequencies using the finite difference method implemented in the \texttt{phonopy} package \cite{Togo:2015} for a $3\times3\times3$ supercell.
For these simulations, the Brillouin zone is sampled by a $3\times3\times3$ MP $\mathbf{k}$-point grid, which leads to phonon frequencies converged to within less than $0.2$ meV.
These phonon calculations are implemented using noncollinear magnetism and include spin-orbit coupling.

\begin{figure}
\includegraphics[width=0.98\columnwidth]{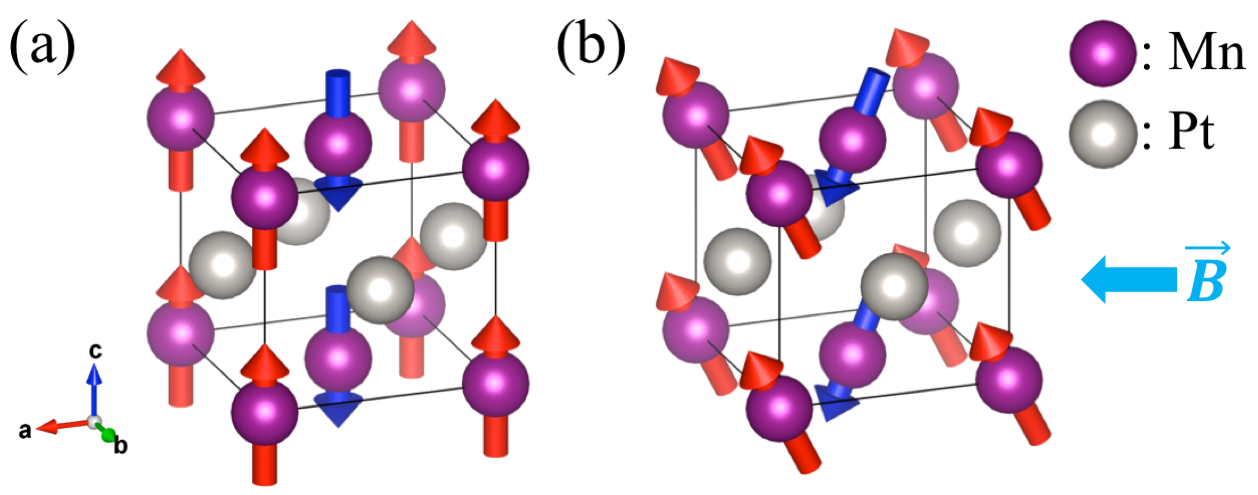}
\caption{\label{fig:strc}(Color online.)
Chemical and magnetic structure of MnPt in the (a) ground state and (b) spin-tilted state under an external magnetic field along $a$-axis direction.
Manganese atoms are purple and platinum atoms are gray.
Red and blue arrows represent antiparallel magnetic moments. 
The magnetic unit cell (shown above) comprises of two chemical unit cells.
}
\end{figure}

We compute the exchange coefficients for antiferromagnetic L1$_{0}$-type MnPt using the spin polarized relativistic Korringa-Kohn-Rostoker (\texttt{SPR-KKR}) code \cite{Ebert:2011}.
The electronic ground state is computed within the KKR formalism, based on the fully relaxed atomic structure and using DFT-PBE \cite{Perdew:1997} as described above.
The Brillouin zone is sampled with 1000 randomly selected $\mathbf{k}$ points, leading to total energies converged within $0.01$ meV/atom.
Isotropic exchange coupling coefficients of a Heisenberg model,
\begin{equation}
\label{eq:exchange}
\mathcal{H}_{ex} = - \sum_{i \neq j}J_{ij}e_{i}e_{j},
\end{equation}
are then computed using Lichtenstein's approach within the \texttt{SPR-KKR} code \cite{Liechtenstein:1984}.
Here $\mathcal{H}_{ex}$ is the exchange Hamiltonian and $J_{ij}$ are exchange coupling parameters for all magnetic moments of atoms $i$ and $j$ and orientations $e_i$ and $e_j$, within an interaction distance $d/a=4.0$, where $a$ is the lattice parameter along the $a$ axis (see Fig.\ \ref{fig:strc}).

Subsequently, we compute magnon dispersion curves within linear spin-wave theory\cite{Toth:2015} from the spin Hamiltonian
\begin{equation}
\label{eq:spinW-Hamil}
\mathcal{H} = \sum_{i,j}\mathbf{S}_{i}^\intercal\mathbf{J}_{ij}\mathbf{S}_{j} + \sum_{i}\mathbf{S}_{i}^\intercal\mathbf{A}_{i}\mathbf{S}_{i},
\end{equation}
that accounts for exchange and anisotropy interactions.
Here $\mathbf{S}_{i}$ is the $3\times1$ spin vector operator, $\mathbf{J}_{ij}$ is the $3\times3$ exchange coupling matrix between spins at sites $i$ and $j$, and $\mathbf{A}_{i}$ is the $3\times3$ anisotropy matrix.
The diagonal components of $\mathbf{J}_{ij}$ can be described by the isotropic exchange coupling parameters in Eq.\,\eqref{eq:exchange}, while the off-diagonal components are Dzyaloshinskii-Moriya exchange parameters.
Due to the inversion symmetry of antiferromagnetic L1$_{0}$-type MnPt, the Dzyaloshinskii-Moriya interaction \cite{Moriya:1960} and, hence, these  off-diagonal components vanish.
$\mathbf{A}_{i}$ represents the anisotropy energy with two-fold symmetry, which follows from the magnetocrystalline anisotropy energy computed within DFT, including spin-orbit interaction and magnetic dipole-dipole interaction (see Sec.\,\ref{sec:grnd-aniso}).
For MnPt with uniaxial magnetism, all components of $\mathbf{A}_{i}$ vanish except for the (3,3) component, which is equal to $(K_{1}+K_{2})/n$, where $n=2$ is the total number of magnetic moments in the magnetic unit cell.
Subsequently, we use the \texttt{SpinW} code \cite{Toth:2015} to compute the magnon dispersion in \textbf{q}-space from the diagonalization of the Fourier transformed spin Hamiltonian.

Finally, we compute the N\'{e}el temperature using a Monte Carlo (MC) method to solve the stochastic Landau-Lifshitz-Gilbert (LLG) equation \cite{Eriksson:2017},
\begin{equation}
\label{eq:StochasticLLG}
\begin{split}
\frac{d\mathbf{m}_{i}}{dt} = &- \gamma_{L}\mathbf{m}_{i} \times (\mathbf{B}_{i}+\mathbf{B}_{i}^\mathrm{fl})\\ 
&- \gamma_{L} \frac{\alpha}{m_{i}} \mathbf{m}_{i} \times \left[ \mathbf{m}_{i} \times (\mathbf{B}_{i}+\mathbf{B}_{i}^\mathrm{fl}) \right],
\end{split}
\end{equation}
where $\mathbf{m}_{i}$ is the magnetic moment at site $i$ consistent with our DFT simulations and $\gamma_{L}=\gamma / (1+\alpha^2)$ is the renormalized gyromagnetic ratio.
$\gamma$ is a gyromagnetic ratio and we use the default value of an isotropic Gilbert damping constant $\alpha$=0.1 implemented in \texttt{UppASD}\cite{Eriksson:2017}, which does not affect our results, since we keep the temperature fixed using a heat bath and, thus, there is no damping.
$\mathbf{B}_{i}$ is the effective magnetic field as the derivative of the spin Hamiltonian, including exchange, anisotropy, and magnetic dipolar interactions, with respect to $\mathbf{m}_{i}$ at magnetic site $i$.
The magnetic temperature is included as a fluctuating magnetic field $\mathbf{B}_{i}^\mathrm{fl}$ based on the central limit theorem, using a Gaussian distribution with zero average and temperature dependent variance \cite{Eriksson:2017}.
The magnetic structure at finite temperature is then calculated from Eq.\ \eqref{eq:StochasticLLG} using a $15\times15\times15$ supercell and the MC approach implemented in the \texttt{UppASD} package \cite{Eriksson:2017}.

\section{\label{sec:grnd}Ground State Properties}

\subsection{\label{sec:grnd-struc}Atomic structure and magnetic configuration}
Antiferromagnetic L1$_{0}$-type MnPt crystallizes in a tetragonal uniaxial structure with a chemical space group of $P4/mmm$ (No.\ 123) and magnetic space group of $C_{P}m'm'm$.
Mn and Pt atoms occupy alternating layers along the $c$ axis, which induces the tetragonal structure (see Fig.\,\ref{fig:strc}).
We first compute fully relaxed lattice parameters using DFT and obtain $a$=3.97\,\AA\ and $c$=3.71\,\AA.
These deviate by less than 1.5\,\% from experimental measurements of $a$=4.00\,\AA\ and $c$=3.67\,\AA\ by Kren \textit{et al.}\ \cite{Kren:1968} and are in even better agreement with another DFT-PBE study by Wang \emph{et al.}\cite{Wang:2013}, reporting $a$=3.98\,\AA\ and $c$=3.72\,\AA.
The collinear antiferromagnetic structure is described by a uniaxial magnetic unit cell with up and down magnetic sites along the [001] easy axis.
Antiparallel magnetic moments are localized on Mn atoms, compensating each other within each layer.
Our DFT results give a sub-lattice magnetization of $\mathbf{M}$=3.7\,$\mu_{\text{B}}$.
The measured value amounts to $\mathbf{M}$=4.3\,$\mu_{\text{B}}$ at room temperature \cite{Kren:1968} and DFT-LDSA results in $\mathbf{M}$=3.6\,$\mu_{\text{B}}$, reported by Umetsu \emph{et al.} \cite{Umetsu:2006}
Our results and those of other experimental and theoretical work are compiled in Table\ \ref{tab:latmag}, from which we conclude that our atomic structure and magnetic configuration is in good agreement with literature data.

\begin{table}
\caption{\label{tab:latmag}
Relaxed lattice parameters (in \AA) along three crystallographic axes and magnetic moments (in $\mu_\mathrm{B}$) of MnPt.
All theoretical results use a spin polarized description without spin-orbit coupling.
}
\begin{tabular}{cccccc}
\hline
MnPt& $a$ & $b$ & $c$ & $\mu_\mathrm{Mn}$ & $\mu_\mathrm{Pt}$ \\
\hline
This work & 3.97&3.97&3.71&3.7&0.0\\
DFT-PBE\ \cite{Wang:2013} &3.98&3.98&3.72&3.7&--\\
DFT-PBE\ \cite{Alsaad:2020} &4.03&4.03&3.69&4.3&0.0\\
DFT-LDSA\ \cite{Lu:2010} &3.99&3.99&3.70&3.8&0.0\\
LMTO-LDSA\ \cite{Umetsu:2006} &--&--&--&3.6&0.0\\
Exp.\ \cite{Kren:1968} &4.00&4.00&3.67&4.3&--\\
Exp.\ \cite{Pearson:1985} &4.002&4.002&3.665&--&--\\
Exp.\ \cite{Severin:1979} &--&--&--&4.0&0.4\\
\hline
\end{tabular}
\end{table}

\subsection{\label{sec:grnd-aniso}Magnetocrystalline anisotropy energy}

The magnetocrystalline anisotropy (MCA) energy of an antiferromagnet originates from contributions due to spin-orbit interaction (SOI) and magnetic dipole-dipole interaction (MDD).
We compute the SOI term using DFT total energies including spin-orbit coupling.
The MDD contribution is a relativistic correction
from transverse electron-electron interactions \cite{Schron:2012} that is not included in the DFT total energy and we describe it here within classical electrodynamics based on the relaxed DFT ground state atomic structure and magnetic moments.
The sum of all MDD interactions in a bulk material is \cite{Ashcroft:1976},
\begin{equation}
\label{eq:MDD}
\begin{split}
E_{\text{MDD}}=-\frac{1}{2}\frac{\mu_{0}}{4\pi}\sum_{i \neq j}\left( \frac{3[\mathbf{m}_{i} \cdot \mathbf{r}_{ij}][\mathbf{m}_{j} \cdot \mathbf{r}_{ij}]}{r_{ij}^{5}}\right.\\
\left. -\frac{[\mathbf{m}_{i} \cdot \mathbf{m}_{j}]}{r_{ij}^{3}}\right),
\end{split}
\end{equation}
where $\mu_{0}$ is the vacuum permeability, $\mathbf{r}_{i}$ is the coordinate of magnetic site $i$, $r_{ij}$ is the distance between two magnetic sites $i$ and $j$, and $\mathbf{m}_{i}$ is the magnetic moment at site $i$.
$E_{\text{MDD}}$ decays as $r_{ij}^{-3}$ and when numerically evaluating Eq.\ \eqref{eq:MDD}, we include interactions within a sphere with a cutoff radius of 50\,\AA. 
This converges $E_{\text{MDD}}$ to within $10^{-9}$ eV/V$_{\textrm{mag}}$, where V$_{\textrm{mag}}$ is the volume of a magnetic unit cell, allowing us to confirm that the MDD anisotropy energy in the \emph{ab}-plane is negligible.

\begin{figure}
\includegraphics[width=0.98\columnwidth]{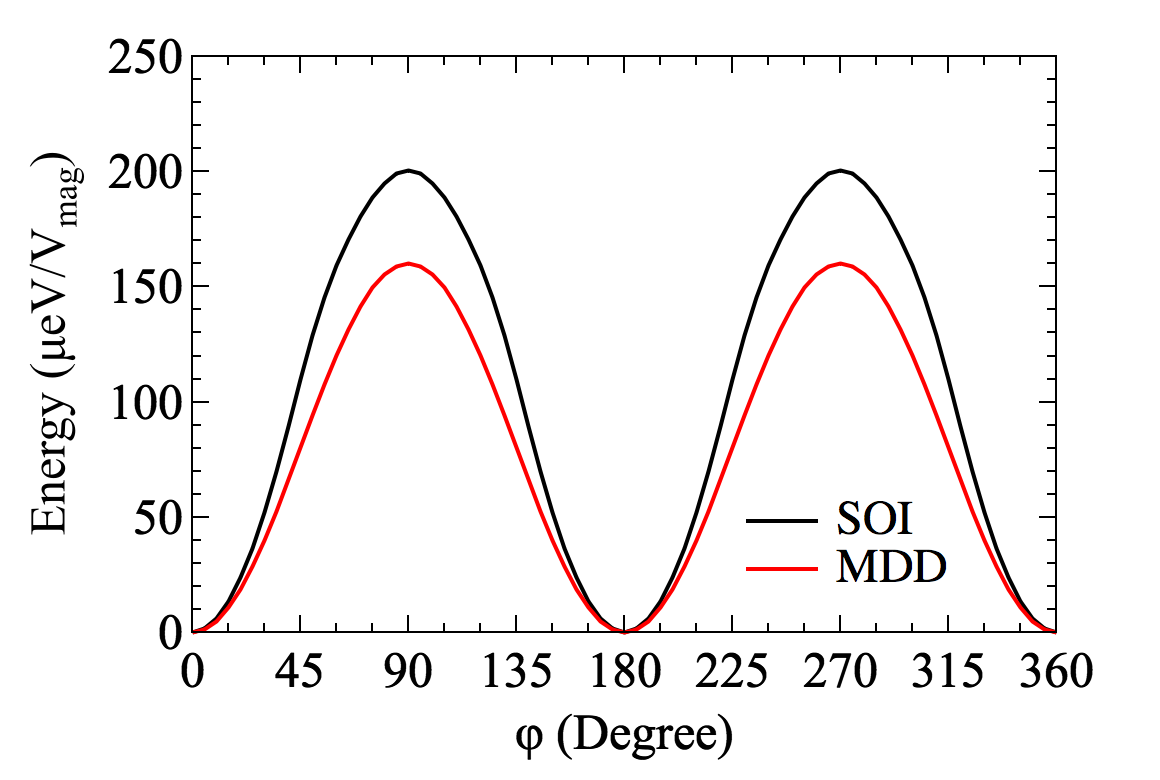}
\caption{\label{fig:aniso}(Color online.)
Magnetocrystalline anisotropy of MnPt shows two-fold periodicity as a function of the tilting angle $\phi$ of the N\'{e}el vector with respect to the $a$ axis with $\theta=22.5^{\circ}$.
Spin-orbit interaction (SOI, black solid line) and magnetic dipole-dipole interaction (MDD, red solid line) contributions are of comparable magnitude.
$V_{\text{mag}}$ is the volume of the magnetic unit cell.
}
\end{figure}

Our MCA energy results for antiferromagnetic L1$_{0}$-type MnPt in Fig.\ \ref{fig:aniso} show two-fold out-of-plane MDD and SOI contributions, confirming uniaxial magnetism.
Using perturbation theory, the MCA energy can be expanded in terms of direction cosines \cite{LANDAU:1984}, which yields for a tetragonal crystal structure \cite{Herak:2009}
\begin{equation}
\label{eq:Aniso}
\frac{E_{\text{MAE}}}{V}=K_{1}\sin^2 \phi + K_{2}\sin^4 \phi + K_{22}\sin^4 \phi \cos(4\theta),
\end{equation}
where $\phi$ describes the angle the N\'{e}el vector forms with the $c$ axis, and $\theta$ is the angle between the $a$ axis and the projection of the N\'{e}el vector to the $ab$ plane.
To study MCA for uniaxial magnetism, we use $\theta$=22.5$^{\circ}$ which corresponds to varying the N\'{e}el vector in the $ac$ plane.
We did not study MCA in the $ab$ plane, because this corresponds to a hard plane.
Fitting our results in Fig.\,\ref{fig:aniso} to Eq.\ \eqref{eq:Aniso} provides us with anisotropy coefficients that we compare to data reported in the literature in Tab.\,\ref{tab:MDD}.
From this we find a significant variation of the results and note that due to their sub-meV magnitude, MCA calculations are very sensitive to details of the computational approach.
In particular, the description of exchange and correlation, lattice parameters, and numerical parameters such as Brillouin zone sampling and plane-wave cutoff energy affect the results and likely explain the range of values reported in the literature.
Here we converge all numerical parameters and estimate our remaining error bar to be about 22\,\%.
In addition, our data illustrates that the MDD contribution to the total MCA energy of $K_{1}=K_{1}^{\mathrm{SOI}}+K_{1}^{\mathrm{MDD}}$=1.07$\times 10^6$ J/m$^3$ ($=391$ $\mu$eV/$V_{\text{mag}}$) is as large as 68\,\% of the SOI contribution and, hence, not negligible.
In antiferromagnetic MnPt the MDD contribution is more important than in antiferromagnetic Fe$_2$As, where we found it to be about 50\,\% of the SOI term \cite{Yang:2020}.

\begin{table}
\caption{\label{tab:MDD}
Magnetocrystalline anisotropy coefficients $K_{1}^{\mathrm{SOI}}$, $K_{2}^{\mathrm{SOI}}$, $K_{1}^{\mathrm{MDD}}$, and $K_{2}^{\mathrm{MDD}}$.
Other theoretical results only use the first term of Eq.\ \eqref{eq:Aniso} to calculate anisotropy, which can be compared to $K_{1}^{\mathrm{SOI}}$+$K_{2}^{\mathrm{SOI}}$ from our simulations.
}
\begin{tabular}{ccccc}
\hline
[$\mu$eV/V$_{\mathrm{mag}}$]& $K_{1}^{\mathrm{SOI}}$ & $K_{2}^{\mathrm{SOI}}$ & $K_{1}^{\mathrm{MDD}}$ & $K_{2}^{\mathrm{MDD}}$ \\
\hline
This work &230&$-29$&160&0\\
DFT-LSDA\ \cite{Lu:2010} &114&--&--&--\\
DFT-LSDA+U\ \cite{Alsaad:2020} &460&--&--&--\\
LMTO-LSDA-ASA\ \cite{Umetsu:2006} &510&--&--&--\\
GF-LMTO\ \cite{Chang:2018} &100&--&--&--\\
\hline
\end{tabular}
\end{table}

\subsection{\label{sec:grnd-sus}Magnetic susceptibility}

The magnetic susceptibility of a material describes how its total energy responds to a change of the magnetic structure in response to an external magnetic field.
When such a field is applied to an antiferromagnetic material, magnetic moments cant towards the field direction, reducing their antiparallel orientation that is energetically favored in the ground state.
The magnetic susceptibility of antiferromagnets connects tilting to a total energy change via the dependence of the exchange energy on tilting. 

We use DFT to compute the magnetic susceptibility from the total energy change resulting from magnetic moment tilting.
The total energy of the electronic system under an applied external magnetic field is \cite{Kang:2020}
\begin{equation}
\label{eq:Etot}
E_\mathrm{tot}=E_0+a\mathbf{\mu}^{2}-{\boldsymbol\mu}\mathbf{B},
\end{equation}
where $E_{0}$ is the ground state total energy without magnetic field, $a\mathbf{\mu}^{2}$ describes the interaction of tilted magnetic moments via an exchange term in a Heisenberg model, ignoring classical dipole-dipole contributions, and $-{\boldsymbol\mu}\mathbf{B}$ is the Zeeman energy term.
$\mathbf{B}$ is the external magnetic field vector and ${\boldsymbol\mu}$ is the induced net magnetization that arises in the presence of the external field.
For the small tilting studied here, the induced magnetic moments are proportional to $\mu$.
We kept all atomic positions fixed when tilting magnetic moments and found that this affects the resulting susceptibility by less than 0.5\,\%.
The lowest energy under an applied field minimizes Eq.\ \eqref{eq:Etot} and corresponds to $\mathbf{B}=2a{\boldsymbol\mu}$ as discussed in Ref.\ \onlinecite{Kang:2020}.
This yields for the magnetic susceptibility perpendicular to N\'{e}el vector.
\begin{equation}
\label{eq:magsus}
\chi_{v}=\frac{\mu_{0}}{2a-\mu_{0}}.
\end{equation}

\begin{figure}
\includegraphics[width=0.98\columnwidth]{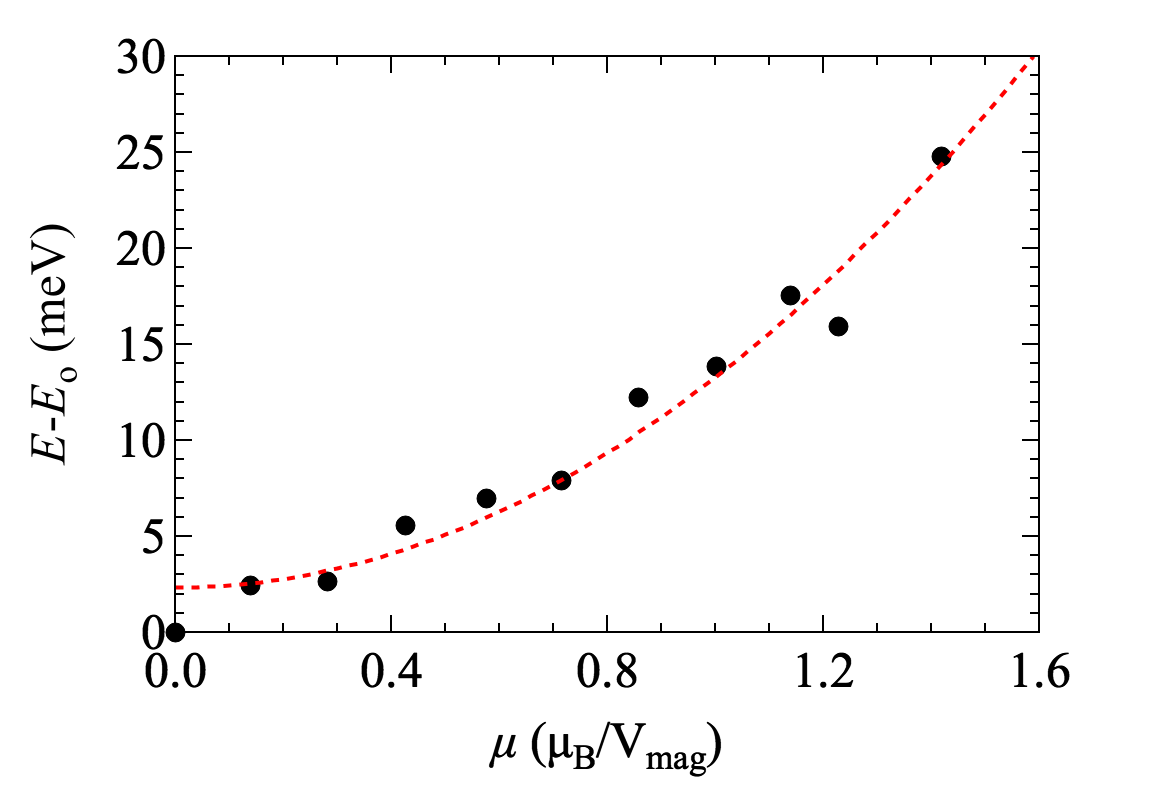}
\caption{\label{fig:magsus}(Color online.)
DFT total energies for different tilting of magnetic moments.
Each point corresponds to a tilting angle between $0^\circ$ and $10^\circ$ with a step size of $1^\circ$.
Red dashed line shows the fit to Eq.\ \eqref{eq:Etot} and the resulting uncertainty for the susceptibility is about 15\,\%.
}
\end{figure}

Here we compute the magnetic susceptibility for an external magnetic field along the [100] crystallographic direction as shown in Fig.\,\ref{fig:strc}(b).
The DFT total energies for magnetic moment tilting between $0^\circ$ and $10^\circ$ degrees (in $1^\circ$ degree increments) in Fig.\,\ref{fig:magsus} show a quadratic dependence on the resulting net magnetization.
A quadratic fit to this curve determines $a$ in Eq.\ \eqref{eq:magsus} and yields a unitless magnetic susceptibility of $5.25\times10^{-4}$, which is in between $4.82\times10^{-4}$ measured at 4.2\,K by Umetsu \emph{et al.}\ \cite{Umetsu:2006} and $6.01\times10^{-4}$ measured at 4.2\,K by Chen \emph{et al.}\ \cite{Chen:1975} on polycrystalline samples.
The magnetic susceptibility of antiferromagnets, including MnPt, is much smaller than that of ferromagnets.
Hence, a large external magnetic field is required to induce a small amount of magnetic moment tilting in antiferromagnets, illustrating the robustness of antiferromagnets against external fields.
For a field oriented parallel to the N\'{e}el vector, i.e.\ the \emph{a}-axis, the magnetic susceptibility would be zero in the limit of zero temperature.

\subsection{\label{sec:grnd-exchng}Exchange coupling coefficients}
\begin{figure}
\includegraphics[width=0.98\columnwidth]{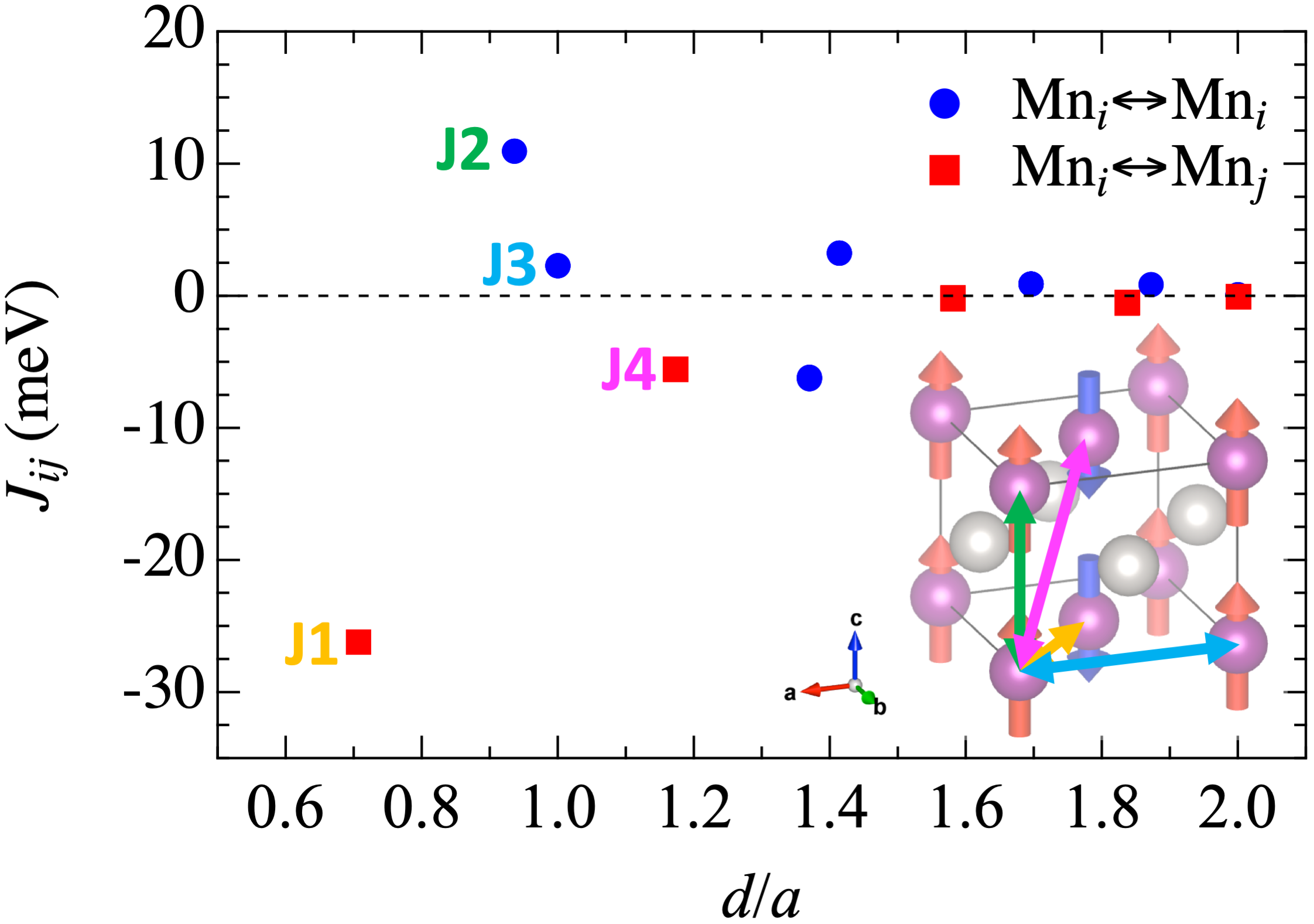}
\caption{\label{fig:magjij}(Color online.)
Exchange coupling coefficients decrease with distance $d$ (in units of the lattice parameter $a$).
Blue circles show interactions of Mn sites with parallel moments, while red squares represent interactions of Mn sites with antiparallel moments.
We include interactions up to eleventh-nearest neighbors for our magnon dispersion calculations.
Colored arrows in the inset figure show the first to fourth neighboring interaction of exchange coupling.
}
\end{figure}

Individual exchange coupling coefficients $J_{ij}$ from a Heisenberg model can be used to explain the magnetic structure of antiferromagnetic L1$_{0}$-type MnPt in detail, in addition to the  description of the collective response of the exchange coupling by the magnetic susceptibility.
We used the (\texttt{SPR-KKR}) code \cite{Ebert:2011} to compute $J_{ij}$ as plotted in Fig.\,\ref{fig:magjij}.
The first coefficient ($J_{ij}$=$-26.2$ meV) indicates antiferromagnetic coupling between any pair of nearest neighbor atoms in the $ab$ plane (see inset of Fig.\ \ref{fig:magjij}).
The sign of the second coefficient ($J_{ij}$=10.9 meV) indicates ferromagnetic coupling between Mn atoms across the Pt layer.
The third coefficient ($J_{ij}$=2.3 meV) corresponds to ferromagnetic coupling in the $ab$ plane, while the fourth interaction ($J_{ij}$=$-5.5$ meV) couples two opposite magnetic moments across the Pt layer with a (1/2, 1/2) shift in the $ab$ plane.
While the Mn in-plane interaction is dominant, the Mn interlayer interaction is non-negligible.
We note that the negative sign of the fifth interaction parameter represents antiferromagnetic coupling of sites with parallel magnetic moments.
While this indicates magnetic frustration, the magnitude of this fifth parameter is too small to affect the magnetic structure.
Finally, we used exchange coefficients up to eleventh-neighbor atoms to compute the magnon dispersion in Sec.\,\ref{sec:enrg-m}.
Using coefficients up to tenth-neighbor atoms changes the magnon dispersion by not more than 0.18 meV, which corresponds to 0.07\,\% of the entire magnon energy scale.

\section{\label{sec:enrg}Energy Dispersion and Heat Capacity}

The energy dispersion of elementary excitations in a material allows to interpret ground state properties, such as magnetocrystalline anisotropy, and excited state properties of materials, including frequency dependent optical spectra or temperature dependent heat capacity.
Here we study the contributions from electrons, phonons, and magnons for antiferromagnetic MnPt and, subsequently, compute the heat capacity contributions from each elementary excitation.
Calculated heat capacity provides direct comparison with experiment, which is used here to validate our computational description.

\begin{figure}
\includegraphics[width=0.98\columnwidth]{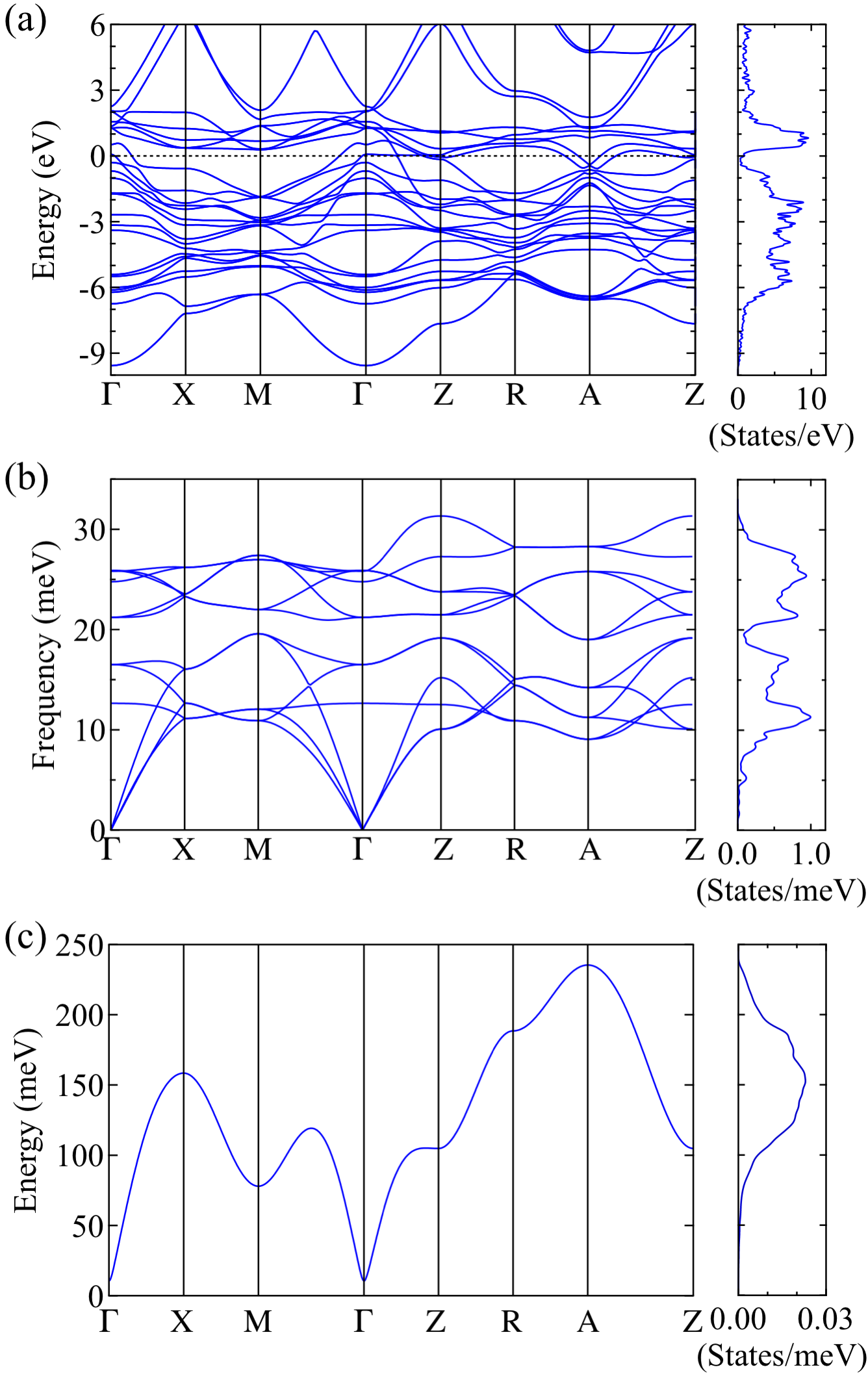}
\caption{\label{fig:disp}(Color online.)
Energy dispersion curves and densities of states (normalized per unit cell) of (a) electrons, (b) phonons, and (c) magnons.
Electronic band structure (a) illustrates the metallic character with low density of states near the Fermi level at $E=0$ eV.
The energy scale of phonons is about one order of magnitude smaller than that of phonons.
Magnon bands from linear spin wave theory show a magnon gap at the $\Gamma$ point.
}
\end{figure}

\subsection{\label{sec:enrg-e}Electronic structure}

Our computed electronic band structure in Fig.\ \ref{fig:disp}(a) accounts for spin-orbit interaction and shows the metallicity of L1$_{0}$-type MnPt.
While bands are crossing at the Fermi level, the density of states itself is very low and exhibits a significant dip within about 0.5 eV.
This is similar to what was reported by Umetsu \emph{et al.}\ from LMTO-LSDA-ASA simulations \cite{Umetsu:2006}, and also agrees with DFT-LSDA simulations by Lu \emph{et al.}\cite{Lu:2010} as well as DFT-PBE by Wang \emph{et al.}\cite{Wang:2013} and Alsaad \emph{et al.}\cite{Alsaad:2020}

We compute the electronic specific heat using the thermodynamic average of the internal energy $U$ at temperature $T$ and the Sommerfeld expansion, leading to \begin{equation}
\label{eq:e-cv}
\gamma_{e}=\frac{\partial U}{\partial T}=\frac{1}{3} \pi^{2} k_{\mathrm{B}}^2 N(E_{\mathrm{F}}),
\end{equation}
where $k_{\mathrm{B}}$ is the Boltzmann constant, and  $N(E_{\mathrm{F}})$ is the density of states at the Fermi level \cite{Umetsu:2006}.
We obtain a value of 0.27\,mJ/(mol K$^2$), which agrees very well with a measured value of 0.26\,mJ/(mol K$^2$) by Umetsu \emph{et al.} \cite{Umetsu:2006} and is slightly smaller than the electronic specific heat of other pure metals.
Their LMTO-LSDA-ASA data \cite{Umetsu:2006} results in 0.33\,mJ/(mol K$^2$) and the DFT-PBE result of Wang \emph{et al.}\cite{Wang:2013} is somewhat lower at 0.13\,mJ/(mol K$^2$).
This difference may originate from our choice of more converged Brillouin zone sampling and plane-wave cutoff that can affect the results of such a small value of the DOS near the Fermi level is computed.
In addition, we note that SOC is included in our DOS simulations, while that seems not to be the case for Refs.\ \onlinecite{Umetsu:2006, Wang:2013}.

\subsection{\label{sec:enrg-p}Phonon dispersion}

Our result for the phonon dispersion in Fig.\ \ref{fig:disp}(b) shows a total of 12 acoustic and optical branches, corresponding to 4 atoms per magnetic unit cell with 3 modes each, with a shallow gap in between at around 18\,meV.
We then use this predicted phonon dispersion to compute the phonon heat capacity from statistical mechanics with the canonical distribution and the harmonic approximation,
\begin{equation}
\label{eq:p-cv}
C_{V}^{\mathrm{phonon}}=\sum_{\mathbf{q}\nu} k_\mathrm{B} [\beta \hbar \omega(\mathbf{q}\nu)]^{2} \frac{\exp{(\beta \hbar \omega(\mathbf{q}\nu))}}{[\exp{(\beta \hbar \omega(\mathbf{q}\nu))}-1]^{2}},
\end{equation}
where $\beta=1/(k_{\mathrm{B}}T)$ \cite{Togo:2015}.
We compute the phonon heat capacity using Eq.\ \eqref{eq:p-cv} and a $30\times30\times30$ $\mathbf{q}$-point grid.
For a linear phonon dispersion near $\Gamma$, a $T^{3}$ dependence follows at low temperatures, as discussed below in Sec.\,\ref{sec:enrg-heat}.

\subsection{\label{sec:enrg-m}Magnon dispersion}

Using linear spin-wave theory, the exchange coefficients from Sec.\,\ref{sec:grnd-exchng}, and the anisotropy coefficients we discussed in Sec.\,\ref{sec:grnd-aniso}, we compute the magnon dispersion shown in Fig.\,\ref{fig:disp}(c).
Since antiferromagnetic MnPt has two magnetic sites, all magnon energy states are doubly degenerate.
We note that the entire magnon energy range reaches up to 250\,meV, which is higher than the band width of phonons of about 30\,meV.
The magnon gap at the $\Gamma$ point is 10.49\,meV (=2.54\,THz).
Our calculated magnon dispersion (see Fig.\ \ref{fig:disp}) includes the anisotropy energy and, hence, it shows an energy gap at $\Gamma$.
Without the anisotropy energy term, this magnon energy gap would disappear and the magnon dispersion would be linear, starting at the $\Gamma$ point.

Next, we use the Kittel formula to compute the lowest magnon frequency $\omega_\mathrm{min}$ from the Landau-Lifshitz equation for $k=0$, which provides an estimate for how fast spin dynamics occurs in MnPt.
For an easy-axis antiferromagnet without external field \cite{Rezende:2019} this leads to,
\begin{equation}
\label{eq:lowfreq}
\omega_\mathrm{min}=\gamma \sqrt{2H_{E}H_{A}+H_{A}^2},
\end{equation}
where $H_{E}$=$m/\chi$ and $H_{A}$=$K/m$ are exchange field and anisotropy field, respectively, $m$ is the magnitude of the sub-lattice magnetization, $\chi$ is the magnetic susceptibility, $K$ is the anisotropy energy coefficient and $\gamma$ is the gyromagnetic ratio ($g\mu_{B}/\hbar$).
Since MnPt has two sites with antiparallel moments, $m$ is identical to the magnetic moment of each of these sites, computed from ground-state DFT.
We use the calculated magnetic susceptibility from Sec.\,\ref{sec:grnd-sus} and the anisotropy energy from Sec.\,\ref{sec:grnd-aniso}.
Our result of $\omega_\mathrm{min}/2\pi$=2.02 THz (8.97 meV) is slightly larger than the magnon gap of 7 meV measured by Hama \emph{et al.}\ at 300 K for vanishing wave vector using inelastic neutron scattering \cite{Hama:2007}.
The small difference between calculated and measured gap may be attributed to a decrease of the anisotropy energy with temperature \cite{Pincus:1959}.
These results also confirm the THz scale of spin dynamics for antiferromagnetic MnPt, which is faster than the GHz scale that is common for ferromagnets, such as 36 and 73\,GHz in ferromagnetic Fe films under dc magnetic fields between 0 to 10\,kOe\cite{Heinrich:1988} and 23.4\,GHz for ferromagnetic garnet films doped with germanium and calcium \cite{He:1985}.
We also note that the spin-flop transition field is closely related to the magnon energy gap, via $H_{\mathrm{sf}}=\sqrt{2H_{E}H_{A}+H_{A}^2}$, resulting in $H_{\mathrm{sf}}=72$\,T.

To compute the magnon heat capacity, we employ the same approach that we used to obtain the magnon dispersion in Fig.\,\ref{fig:disp}(c), to compute the magnon density of states on a $30\times30\times30$ $\mathbf{q}$-point grid.
Since magnons are bosonic, the magnon total energy follows from
\begin{equation}
\label{eq:enrg-m-disc}
E_{\mathrm{magn.}}=\sum_{\mathbf{q}v} \hbar \omega(\mathbf{q}v) \frac{1}{\exp{(\beta \hbar \omega(\mathbf{q}v))}-1}.
\end{equation}
The temperature derivative of this expression leads to the magnon heat capacity
\begin{equation}
\label{eq:cv-m}
C_{V}^{\mathrm{magn.}}=\sum_{\mathbf{q}v} k_\mathrm{B} [\beta \hbar \omega(\mathbf{q}v)]^{2} \frac{\exp{(\beta \hbar \omega(\mathbf{q}v))}}{[\exp{(\beta \hbar \omega(\mathbf{q}v))}-1]^{2}},
\end{equation}
which resembles the expression for the phonon heat capacity, Eq.\ \eqref{eq:p-cv}.
This approach is valid for the low temperature range, the so-called spin-wave region, and our result for the magnon specific heat is shown in Fig.\,\ref{fig:heat}.
At high temperature near the critical temperature, the spin-wave description is no longer valid.
Thus, we describe the critical region near the N\'{e}el temperature using a Monte Carlo approach instead, as discussed in Sec.\,\ref{sec:exct-Tn}.

\subsection{\label{sec:enrg-ins}Inelastic neutron scattering simulation}

\begin{figure*}
\includegraphics[width=0.98\textwidth]{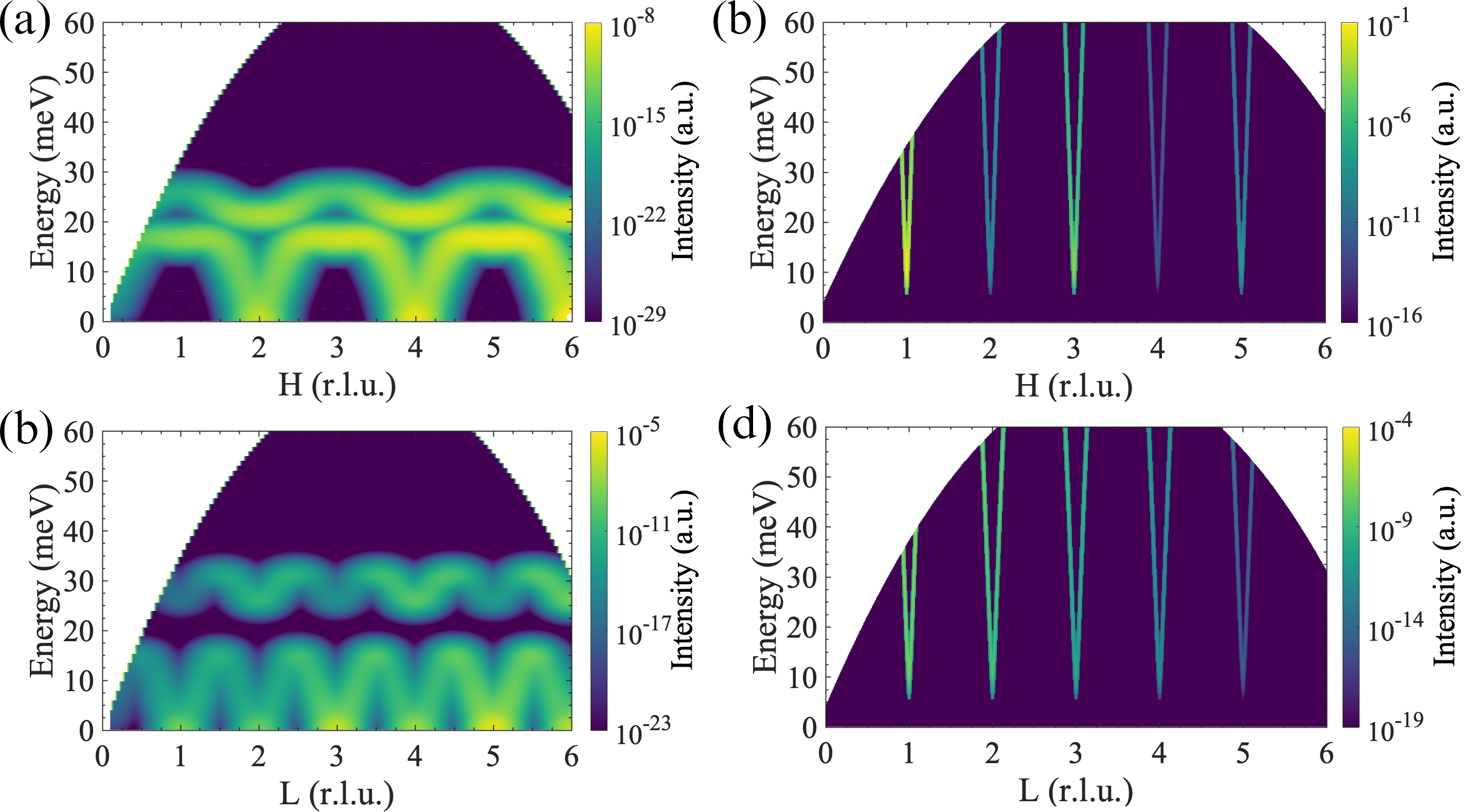}
\caption{\label{fig:ins}(Color online.)
Phonon ((a), (c)) and magnon ((b), (d)) contributions to simulated inelastic neutron scattering along [H 0 0] direction ((a), (b)) and [0 0 L] direction ((c),(d)) in reciprocal space (in reciprocal lattice units, r.l.u).
We use a logarithmic color scale to show the upper 90\,\% of the intensity data.
Magnon curves include a Gaussian broadening of 10 meV and phonon curves are broadened using the \texttt{OCLIMAX} code for a temperature of 5 K.
}
\end{figure*}

To facilitate comparison of our phonon and magnon dispersion data with experiment, we simulate inelastic neutron scattering (INS) intensity using the same instrument parameters as in our previous study on Fe$_2$As \cite{Karigerasi:2020}.
Coherent and incoherent inelastic neutron scattering \cite{Cheng:2019} is added to our phonon results using the \texttt{OCLIMAX} code.
For the magnon contribution, the dynamical spin-spin correlation function is included using the \texttt{SpinW} code \cite{Toth:2015}.
Both simulated INS results are shown in Fig.\,\ref{fig:ins} for the [H00] (symmetrically identical to [0K0]) and [00L] directions commonly studied in experiment.
The phonon form factor is proportional to \textbf{q}$^2$, which explains the intensity increase of the phonon contribution.
Conversely, the magnon contribution weakens with increasing \textbf{q}-vector.
In correspondence with Fig.\,\ref{fig:disp}, all phonon-related signals appear below 40\,meV.
The magnon signals increase as sharp linear lines beyond 40\,meV and, as a result, appear only in the close vicinity of the $\Gamma$ points in Fig.\ \ref{fig:ins}.
The phonon contribution to INS along H in Fig.\,\ref{fig:ins}(a) shows periodicity with every two reciprocal lattice periods, while the signal along L presents the same periodicity as shown in Fig.\,\ref{fig:disp}.
For the magnon signal in Fig.\,\ref{fig:ins}(b) we find alternating intensities for even and odd reciprocal lattice periods, since the magnetic unit cell comprises of two chemical unit cells in the $ab$ plane (see Fig.\,\ref{fig:strc}).

Our analysis shows that phonon and magnon contributions can be clearly distinguished in INS experimental data.
In experiment, the magnon gap energy at $\mathbf{q}=0$ is determined by finding the energy where the INS intensity is at a maximum for $(1 0 0)$ along the H direction in reciprocal space.
The calculated INS data shown in Fig.\,\ref{fig:ins} demonstrates that there is no phonon contribution at this point.
Therefore, the signal clearly originates from magnons, as reported in the INS study of Hema \emph{et al.}\ \cite{Hama:2007}, confirming that the corresponding gap energy is a magnon gap.

\subsection{\label{sec:enrg-heat}Total heat capacity}

\begin{figure}
\includegraphics[width=0.98\columnwidth]{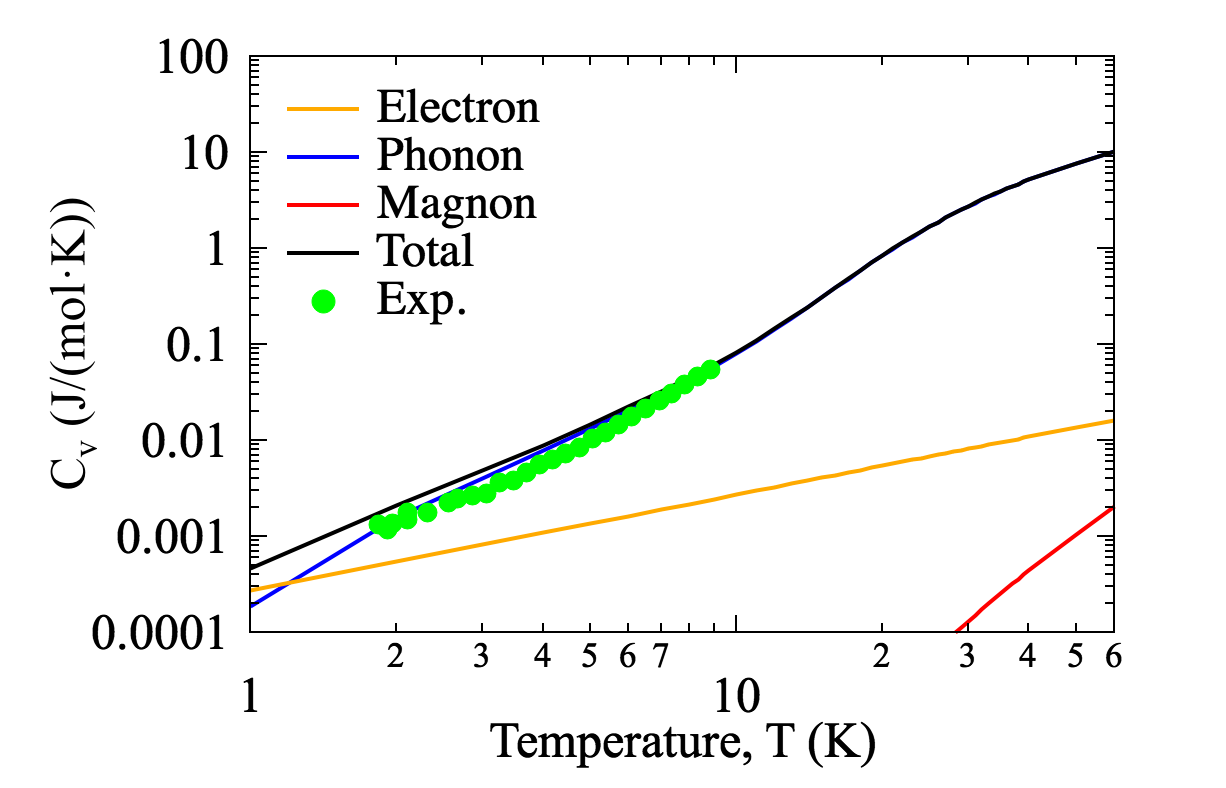}
\caption{\label{fig:heat}(Color online.)
Temperature dependence of electron, phonon, and magnon contributions to the total heat capacity.
Our first-principles results for electron, phonon, and magnon contributions to the total specific heat agree well with experimental data from Ref.\ \onlinecite{Umetsu:2006} at low temperatures.
}
\end{figure}

In Fig.\ \ref{fig:heat} we show the total heat capacity and partition it into electron, phonon, and magnon contributions.
This illustrates that at temperatures below 2\,K, the electronic contribution is dominant and at higher temperatures the phonon contribution takes over with a $T^{3}$ dependence, which is consistent with a linear phonon dispersion near $\Gamma$ \cite{Ashcroft:1976, Kittel:2004}.
The prefactor of the $T^{3}$ term due to phonons is 0.090\,mJ/(mol K$^{4}$).
The onset of the magnon contribution appears at non-zero temperature due to the $\Gamma$ point gap of the magnon dispersion that results from the nonzero anisotropy energy.
The low magnon contribution to the heat capacity at low temperatures further results from the low magnon density of states in the energy range below 50\,meV.
The total number of phonon modes is twelve per four-atom magnetic unit cell, while that of magnons is two per four-atom magnetic unit cell with two magnetic moments.
Finally, the electronic specific heat contributes linearly with $T$, which determines the total heat capacity near 0\,K as shown in Fig.\ \ref{fig:heat}.
We note that our computed total heat capacity in Fig.\ \ref{fig:heat} agrees well with measured results and show that the overall temperature dependence is thus dominated by the \emph{phonon} contribution in the low temperature range.
The lower magnon density of states leads to a lower magnon heat capacity, compared to phonons, see Figs.\ \ref{fig:disp}(b) and (c).

\section{\label{sec:exct}N\'{e}el Temperature and Magneto-Optical Properties}

Excited state properties are popularly used as materials selection criteria, to identify materials well-suited for specific applications, and provide insight into the physics of the AFM since they derive from the electronic band structure including spin-orbit coupling.
In particular, first-principles studies can predict the N{\'{e}}el temperature, which determines thermal stability of the AFM configuration, and magneto-optical effects, that play a role for magnetic characterization.

\subsection{\label{sec:exct-Tn}N\'{e}el temperature}

\begin{figure}
\includegraphics[width=0.98\columnwidth]{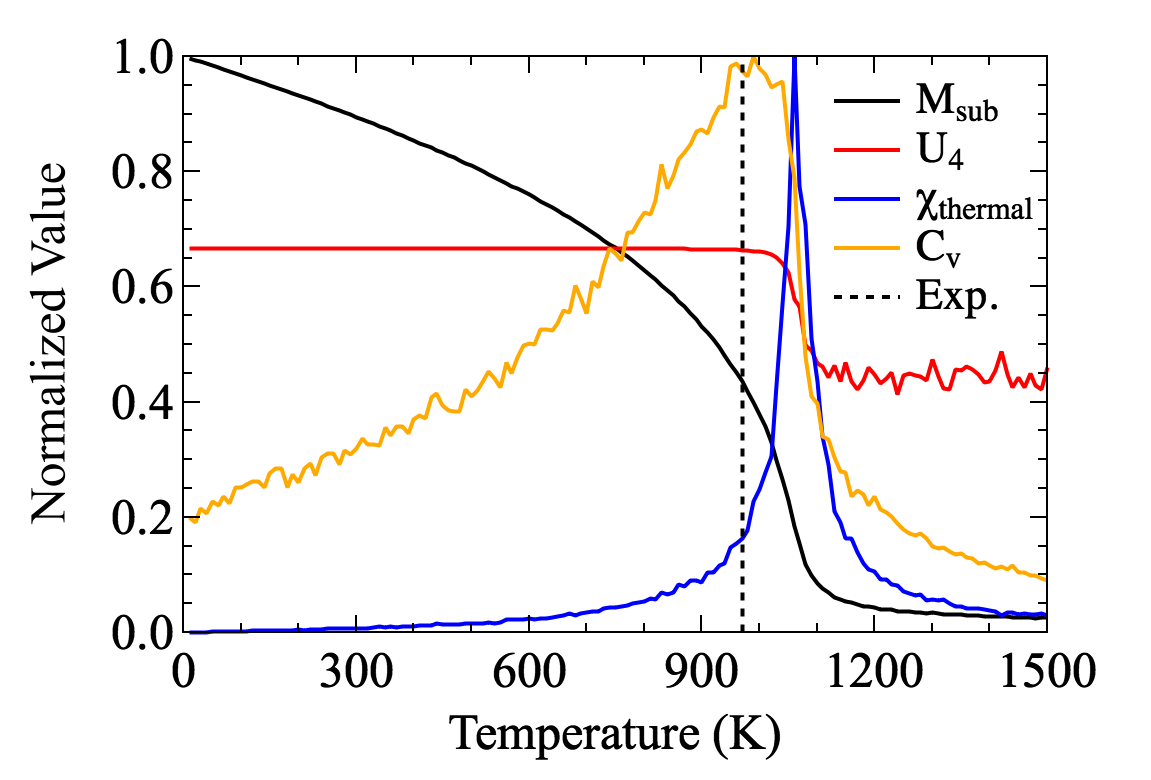}
\caption{\label{fig:thermo}(Color online.)
Temperature dependence of sub-lattice magnetization ($M_{\textrm{sub}}$), isothermal susceptibility ($\chi_{\mathrm{thermal}}$), and heat capacity ($C_{V}$).
These are normalized using the respective maximum values in this temperature range, i.e., the ground-state sub-lattice magnetization for $M_{\textrm{sub}}$, and the peak values at the critical temperature for $\chi_{\mathrm{thermal}}$ and $C_{V}$.
The fourth order Binder cumulant $U_4$ is computed from Eq.\ \eqref{eq:Binder} and shown as red solid line.
}
\end{figure}

We compute the N\'{e}el temperature from thermodynamic observables using a Monte-Carlo solution of the stochastic LLG equation, Eq.\ \eqref{eq:StochasticLLG}, parametrized by our calculated exchange interactions.
We use this approach since, near the critical temperature, the linear-spin wave approach discussed in Sec.\,\ref{sec:enrg-m} is not applicable.
In the MC approach, the average sub-lattice magnetization is typically studied as a function of temperature, and should be zero at the critical temperature.
However, due to finite size effects in our simulations, this transition cannot be easily detected and is not very sharp in Fig.\ \ref{fig:thermo}.

Instead, the Binder cumulant, the isothermal susceptibility, and the specific heat are thermodynamic observables that provide a clearer picture  \cite{Eriksson:2017}.
The fourth-order Binder cumulant $U_4$ was specifically developed to correct the finite size problem for second-order phase transitions \cite{Binder:1981},
\begin{equation}
\label{eq:Binder}
U_{4} = 1 - \frac{\left \langle m^{4} \right \rangle}{3\left \langle m^{2} \right \rangle^{2}},
\end{equation}
where $m$ is the magnitude of the sub-lattice magnetization.
In this work, $m$ is identical to the magnetic moment at each magnetic site because MnPt has two sites with antiparallel moments.
The value of the cumulant changes at the N\'{e}el temperature from $U_L \approx 0.444$ for $T>T_{N}$ to $U_L \approx 0.667$ for $T<T_{N}$.
From this, we compute a transition temperature of around 1070\,K.
The isothermal susceptibility of a sub-lattice susceptibility
\begin{equation}
\label{eq:isothermal}
\chi_\mathrm{thermal} = \frac{\left \langle m^{2} \right \rangle - \left \langle m \right \rangle^{2}}{k_{B}T}
\end{equation}
is another thermodynamic observable which describes the response of the magnetization to temperature, where $m$ is the sub-lattice magnetization, $k_{B}$ is the Boltzmann constant, and $T$ is temperature.
It peaks at around 1060\,K, which is close to the value from the Binder cumulant.
Lastly, the heat capacity
\begin{equation}
\label{eq:heat}
C_{V}=\frac{\left \langle E^{2} \right \rangle - \left \langle E \right \rangle^{2}}{k_{B}T^{2}}
\end{equation}
can be computed from the variance of the MC total energy $E$ and leads to a peak of the specific heat around 990\,K.
These results agree well with measured N\'{e}el temperatures of 975 K \cite{Kren:1968} and 970 K \cite{Umetsu:2006}.

In addition, the N\'{e}el temperature can be computed without the MC approach, via integration of the adiabatic magnon dispersion in Fig.\,\ref{fig:disp}(c).
Two methods are commonly used in the literature \cite{Essenberger2011},
one is based on the mean-field approximation (MFA)
\begin{equation}
\label{eq:Tn-MFA}
k_\mathrm{B}T_{N}^{\mathrm{MFA}} = \frac{M}{3} \left [ \frac{1}{N} \sum_{\mathbf{q}=\mathbf{0}}^\mathrm{BZ} \omega(\mathbf{q}) \right ],
\end{equation}
and another on the random phase approximation (RPA)
\begin{equation}
\label{eq:Tn-RPA}
k_\mathrm{B}T_{N}^{\mathrm{RPA}} = \frac{M}{3} \left [N \sum_{\mathbf{q}=\mathbf{0}}^\mathrm{BZ} \frac{1}{\omega(\mathbf{q})} \right ]^{-1}.
\end{equation}
In both expressions $m$ stands for the sub-lattice magnetization and $N$ is the total number of magnon energies sampled by the $\mathbf{q}$ point grid.
Here we use a $30\times30\times30$ $\mathbf{q}$-point grid to evaluate these expressions and obtain $T_N^{\mathrm{MFA}}$=1250 K and $T_N^{\mathrm{RPA}}$=1190 K.
Both values are slightly larger than measured values of 975 K \cite{Kren:1968} and 970 K \cite{Umetsu:2006}, or another DFT result, using exchange coefficients, of 989 K \cite{Alsaad:2020}.
A similar overestimation of the experimental result on the order of 25\,\% by this approach is also reported e.g.\ for ferromagnetic bcc Fe and antiferromagnetic NiO.
The MC approach shows better agreement with experiment because the methods based on magnon dispersion assume the spin-wave regime, which is only appropriate at low temperatures relative to the N\'{e}el temperature.

\subsection{\label{sec:exct-optics}Optical response}

In this work we use the Kohn-Sham electronic structure, including spin-orbit coupling effect, to compute optical spectra of MnPt.
First, we compute the imaginary part of the interband contribution to the complex, frequency-dependent dielectric tensor \cite{Gajdos:2006} using 
\begin{equation}
\label{eq:optic-inter}
\begin{split}
\varepsilon_{\alpha \beta}^{(2)}=\frac{4 \pi^{2} e^{2}}{\Omega} \lim_{q\rightarrow 0} & \frac{1}{q^{2}} \sum_{c,v,\mathbf{k}} 2 \omega_{\mathbf{k}} \delta (\epsilon_{c\mathbf{k}}-\epsilon_{v\mathbf{k}}-\omega)\times \\
&\times \left \langle u_{c\mathbf{k}+e_{\alpha}q}|u_{v\mathbf{k}} \right \rangle \left \langle u_{c\mathbf{k}+e_{\alpha}q}|u_{v\mathbf{k}} \right \rangle ^{*},
\end{split}
\end{equation}
where $\alpha$ and $\beta$ are Cartesian indices, $\Omega$ is the unit cell volume, $w_{k}$ is the symmetry weight of each $\mathbf{k}$-point, $c$ and $v$ index conduction and valence bands, $\epsilon_{c\mathbf{k}}$ and $\epsilon_{v\mathbf{k}}$ are Kohn-Sham eigenvalues, and $u_{c\mathbf{k}}$ and $u_{v\mathbf{k}}$ are the cell periodic part of the Kohn-Sham orbitals.
The real part, $\varepsilon_{\alpha \beta}^{(1)}$, follows from the imaginary part, $\varepsilon_{\alpha \beta}^{(2)}$, via Kramers-Kronig transformation.

Since antiferromagnetic MnPt is metallic (see Sec.\,\ref{sec:enrg-e}), intraband contributions to the dielectric tensor need to be included, in addition to the interband contributions in Eq.\ \eqref{eq:optic-inter}.
We use the Drude equation,
\begin{equation}
\label{eq:optic-intra}
\varepsilon(\omega) = - \frac{\omega_{p}^{2}}{\omega^{2}+i\omega\Gamma_{D}},
\end{equation}
where $\omega_{p}$ is the plasma frequency and $\Gamma_{D}$ is the line width originating from the finite electron lifetime.
We compute the plasma frequency from our Kohn-Sham electronic structure \cite{Harl:2007}, using 
\begin{equation}
\label{eq:plasma}
\omega^{2}_{p,\alpha\beta} = \frac{4 \pi e^{2}}{\Omega \hbar^{2}} \sum_{n,\mathbf{k}} 2 g_{\mathbf{k}} \frac{\partial f(\epsilon_{n\mathbf{k}})}{\partial \epsilon} \left ( \mathbf{e}_{\alpha} \frac{\partial \epsilon_{n\mathbf{k}}}{\partial \mathbf{k}} \right ) \left ( \mathbf{e}_{\beta} \frac{\partial \epsilon_{n\mathbf{k}}}{\partial \mathbf{k}} \right ).
\end{equation}
Various scattering mechanisms affect the electron lifetime, including electron-electron and electron-phonon scattering, and this value is challenging to compute from first principles \cite{Bernardi:2014,Bernardi:2015}.
Instead, here we use the electric resistivity of 21\,$\mu\Omega\cdot$cm at 300\,K measured by Umetsu \emph{et al.}\ \cite{Umetsu:2006} and our value for the plasma frequency of $\omega_P$=5.29\,eV to estimate the electron scattering time quasi-classically as $\Gamma_{D}$=$1/(\epsilon_{0}\omega_{p}^2\rho)$=7.44\,fs, which corresponds to a lifetime broadening of 0.56\,eV.

\begin{figure}
\includegraphics[width=0.98\columnwidth]{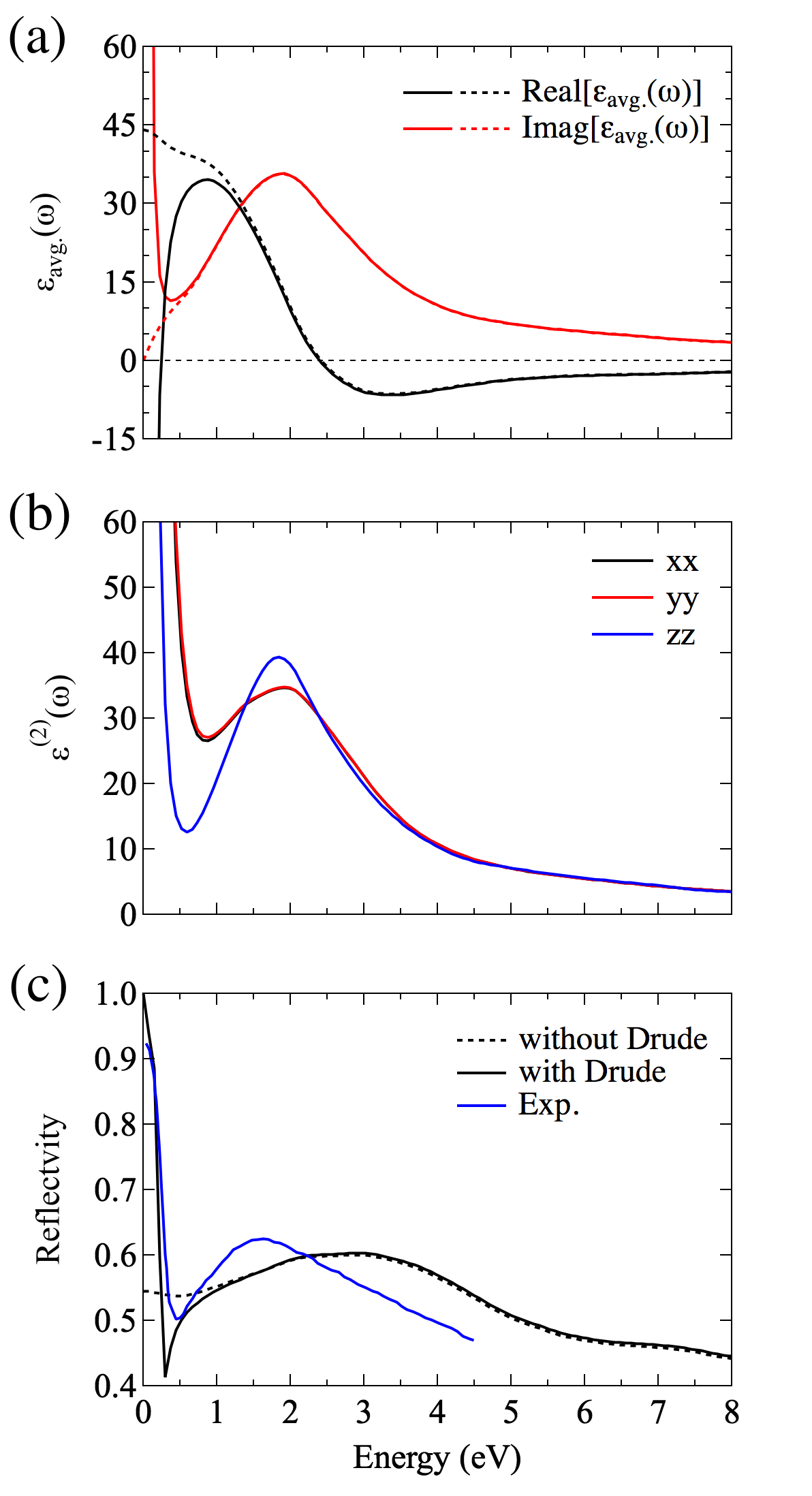}
\caption{\label{fig:optics}(Color online.)
(a) Real (black) and imaginary (red) part of the complex dielectric tensor, averaged over the Cartesian components.
(b) Imaginary part of the three diagonal elements of the complex dielectric tensor.
(c) Reflectivity of antiferromagnetic MnPt, experimental results are from Kubota \emph{et al.}\ \cite{Kubota:2007}
Solid and dashed lines show our simulation results with and without intraband Drude contribution, respectively.
}
\end{figure}

Comparing our calculated frequency dependent dielectric functions with and without Drude contribution in Fig.\ \ref{fig:optics}(a) illustrates that the intraband contribution predominantly affects the low energy range below 1\,eV.
In particular, the interplay of intra- and interband contribution leads to a valley of $\varepsilon_{\alpha \beta}^{(2)}$ for a photon energy of about 0.5 eV.
Anisotropic dielectric functions along three crystallographic axes directions are shown in Fig.\,\ref{fig:optics} (b).
Due to the tetragonal crystal structure of this material, the $xx$ and $yy$ components of the dielectric tensor show identical spectra, and the $zz$ component differs.

We also compute the reflectivity \cite{Fox:2010} from
\begin{equation}
\label{eq:optic-rflc}
R=\left | \frac{\tilde{n}-1}{\tilde{n}+1} \right |^{2} = \left | \frac{\sqrt{\tilde{\epsilon}}-1}{\sqrt{\tilde{\epsilon}}+1} \right |^{2},
\end{equation}
where $\tilde{n}$ is the complex refractive index which is the square root of the averaged diagonal components of complex relative dielectric constant, $\tilde{n}^{2}=\tilde{\epsilon}$.
The resulting reflectivity spectra are plotted in Fig.\,\ref{fig:optics}(b) and compared to experimental data from Kubota \emph{et al.}\ \cite{Kubota:2007}
We find that the overall spectrum agrees well between experiment and simulation, but the position of the low-energy reflectivity mininum differs between experiment (0.45 eV) and simulation (0.29 eV).
Also the position of a broad higher energy reflectivity peak disagrees between 1.63 eV (experiment) and 2.95 eV (simulation).
In addition, the comparison of the reflectivity with and without Drude contribution confirms that the high reflectivity at low photon energies originates from intraband transitions.

\subsection{\label{sec:exct-moke}Linear Magneto-Optical Kerr Effect}

While most collinear antiferromagnets do not show linear magneto-optical effects \cite{Smejkal:2020}, it is possible to generate such signals using spin precession \cite{Kimel:2009} or external stimulation e.g.\ via an electric field \cite{Sivadas:2016}.
Applying an external magnetic field also can break the $C_{P}m'm'm$ magnetic space group symmetry of antiferromagnetic MnPt, leading to non-zero linear magneto-optical Kerr effect.
Here, we introduce such a field perpendicular to N\'{e}el vector (\emph{a}-axis direction) by tilting the magnetic moments between $0^{\circ}$ and $3^{\circ}$ in steps of $1^{\circ}$, inducing a small net magnetization, see Fig.\,\ref{fig:strc}(b).
We then follow Ref.\ \onlinecite{Sangalli:2012} and compute the frequency-dependent polar magneto-optical Kerr effect (PMOKE) using
\begin{equation}
\label{eq:PMOKE}
\Psi_{K}(\omega)=\theta_{K}(\omega)+i\gamma_{K}(\omega)=\frac{-\epsilon_{xy}}{(\epsilon_{xx}-1)\sqrt{\epsilon_{xx}}}.
\end{equation}
All calculations include the Drude contribution, assuming the constant electron lifetime discussed in Sec.\,\ref{sec:exct-optics}.

\begin{figure}
\includegraphics[width=0.98\columnwidth]{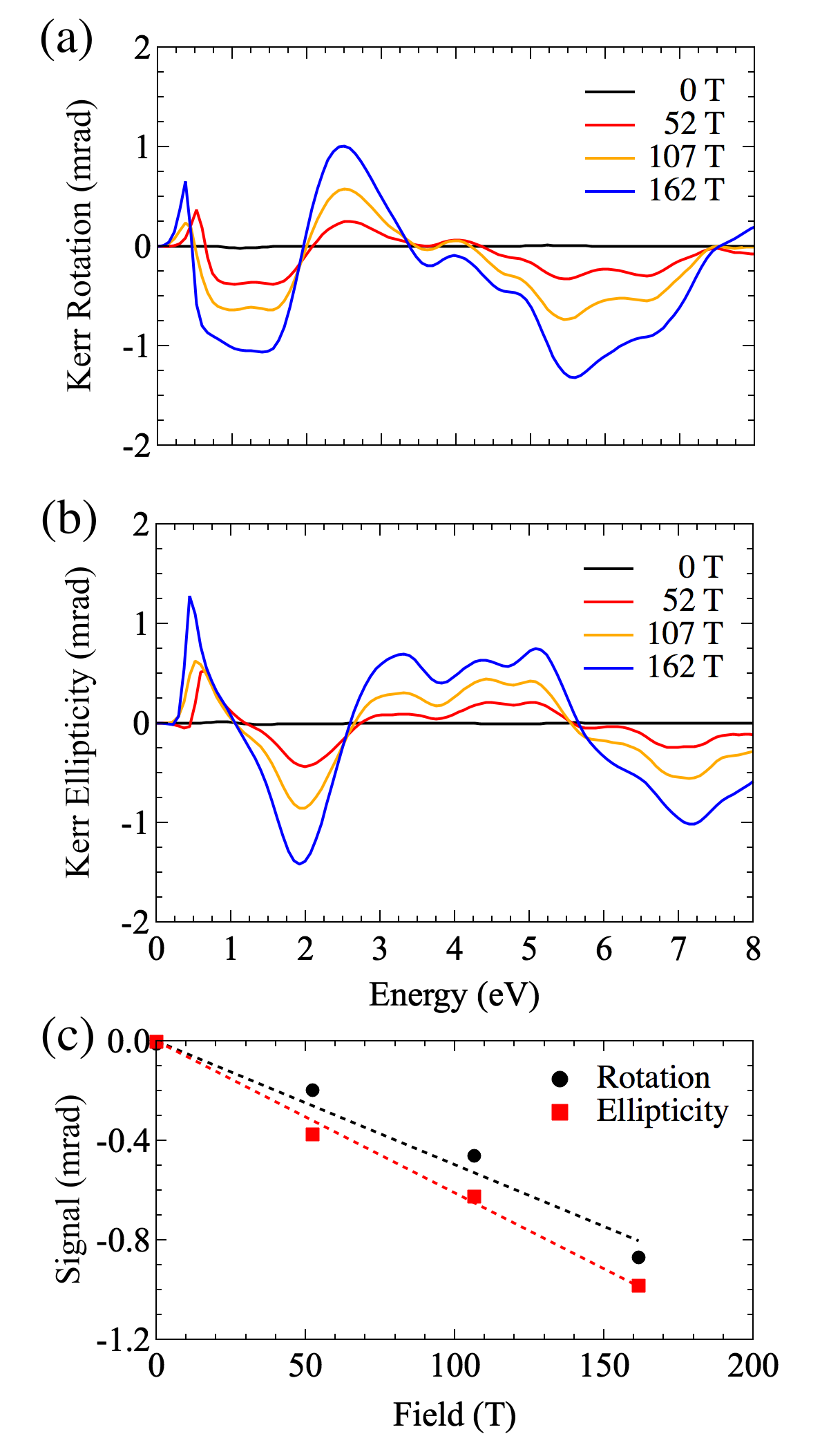}
\caption{\label{fig:optB}(Color online.)
Optical polar magneto-optical Kerr (a) rotation and (b) ellipticity spectra for different external magnetic fields.
Maxima of Kerr rotation and ellipticity occur at 1.40 eV and 1.91 eV, respectively.
In (c) the linear dependence of the Kerr signals on the magnetic field is shown for a wave length of 785\,nm ($=1.58$\,eV).
}
\end{figure}

Figure \ref{fig:optB}(a) and (b) show the resulting PMOKE rotation and ellipticity spectra.
The field strength is computed from the tilting angle using the magnetic susceptibility discussed in Sec.\,\ref{sec:grnd-sus}. 
Due to the small magnetic susceptibility compared to ferromagnetic materials, tilting angles of $1^{\circ}$ correspond to an external magnetic field of 52 T for antiferromagnetic MnPt.
From Fig.\ \ref{fig:optB} we find maximum Kerr rotation and ellipticity in the visible spectral range near 1.40\,eV and 1.91\,eV, respectively.

Our results also show that the interband PMOKE signal, at energies larger than about 1 eV, is approximately proportional to the external magnetic field.
This can be understood from a Taylor expansion of the dielectric function with respect to net magnetization \cite{Siddiqui:2020}.
For small tilting angles and small net magnetization the proportionality of linear MOKE with $B$ is valid, i.e.\ $\Delta\varepsilon \propto \boldsymbol{\mu} \propto \boldsymbol{B}$.
The approximately linear dependence of the Kerr signals on the magnetic fields studies in this work is explicitly shown for a laser wave length of 785 nm in Fig.\,\ref{fig:optB}(c).
From interpolating the linear fit to this data to a magnetic field of 1 T, we find a Kerr rotation and ellipticity of $-6.1$\,$\mu$rad and $-5.0$\,$\mu$rad, respectively.

\begin{figure}
\includegraphics[width=0.98\columnwidth]{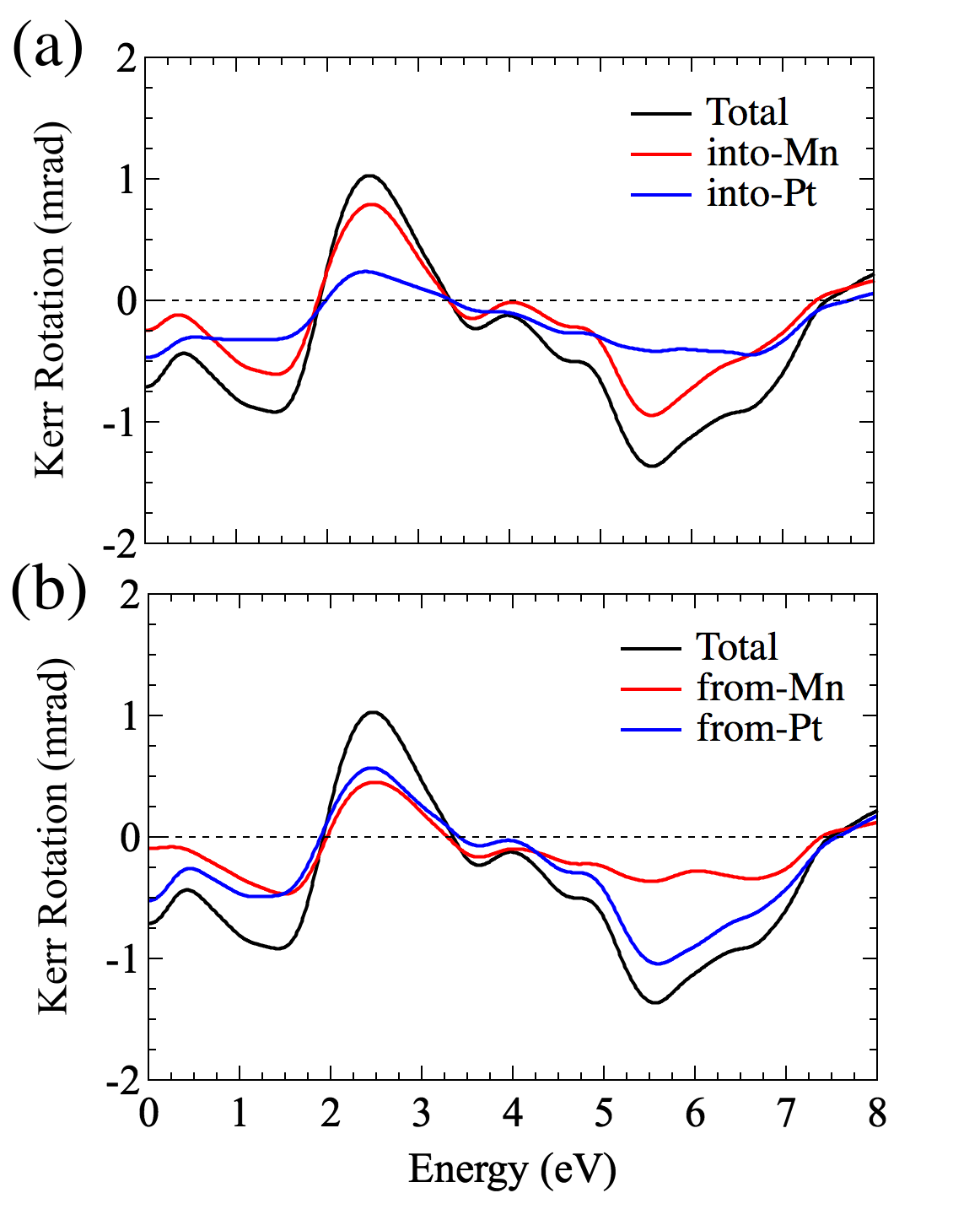}
\caption{\label{fig:spec-decomp}(Color online.)
Projected element orbital decomposition of PMOKE Kerr rotation at 3$^{\circ}$ tilting angle contributed by (a) transitions from all valence states to conduction states of specific atom element and (b) transitions from valence states of specific atom element to all conduction states. Spectrum in this figure does not include the intraband transition contribution. Black solid line shows total Kerr rotation spectrum from interband transitions.
}
\end{figure}

To determine the origin of features in the Kerr rotation spectrum, we decompose it according to contributions from valence and conduction electrons of Mn and Pt states using the scheme described in Refs.\ \onlinecite{Leveillee:2018,Kang:2020}.
The data in Fig.\ \ref{fig:spec-decomp}(a) illustrates that peaks at 1.42 eV, 2.46 eV, and 5.57 eV feature large contributions due to transitions into empty Mn states.
Transition originating in Mn valence states contribute about the same to the PMOKE spectrum across the entire spectral range as those from Pt valence states, except for the peak at 5.57 eV, that is dominated by Pt valence states.
Furthermore, our orbital decomposition concludes that transitions among $d$ orbitals are the main source of the PMOKE spectrum, which is consistent with the majority of states near the Fermi level exhibiting $d$ orbital characters (see Fig.\,S1 in the supplementary material).

\section{\label{sec:cncl}Conclusions}

As the interest in antiferromagnetic spintronics increases, fundamental properties of antiferromagnetic metals, and their accurate prediction from first principles become increasingly important.
Here we report a comprehensive first-principles computational study of antiferromagnetic L1$_{0}$ type MnPt.
For the lattice geometry, and electronic and magnetic structure we find very good agreement with earlier experimental and theoretical results.
Similarly, our prediction of the magnetic susceptibility agrees well with experimental data.
We then compute the previously unknown exchange coupling coefficients and discuss how these explain the ground-state magnetic structure.
Using these coefficients, we predict the magnon dispersion of MnPt, including the lowest magnon frequency of $8.97$\,meV, which is critical for a deep understanding of fundamental limits of the time scale of spin dynamics.
The corresponding gap at the $\Gamma$-point of the magnon dispersion agrees well with the lowest magnon frequency computed using spin-wave theory and we also find very good agreement with an experimentally reported value.

Having established the accuracy of our first-principles description, we proceed to compute electron, phonon, and magnon dispersion data that we use to derive the individual contributions to the heat capacity of MnPt.
We unambiguously show that the temperature dependence of the heat capacity is dominated by phonon contributions at low temperatures, and the magnon contribution remains small, owing to the sizable magnon gap and the low magnon density of states.
Using our data, we individually predict phonon and magnon contributions to inelastic neutron scattering, which will facilitate identification of each contribution in experiment.
Phonon inelastic neutron scattering shows a periodicity over two reciprocal lattice units along the $H$ direction, while the magnon signal presents alternating intensities with a periodicity of one reciprocal lattice unit.
This is because two magnetic Mn atoms are placed along [100] when viewed along [010].
The broader energy range and characteristic linear magnon dispersion curves that originate from every reciprocal lattice unit allow distinguishing phonons and magnons experimentally.

In order to explore the stability of the magnetic ordering and the possibility of reorienting the N\'{e}el vector, we compute the magnetocrystalline anisotropy energy, and find confirmation of the uniaxial antiferromagnetic structure of the material.
We explicitly include a classical contribution to this energy that accounts for magnetic dipole interactions and previously was ignored for antiferromagnets.
Our simulations provide clear evidence for the importance of this contribution to the $K_{1}$ anisotropy coefficient, as it amounts to about 2/3 of the commonly studied term due to the spin-orbit interaction.
In addition, we employed the Monte Carlo method with our calculated exchange and anisotropy coefficients to
compute three thermodynamic observables from atomistic spin dynamics, from which we estimate the N\'{e}el temperature to be 990--1070 K, which is within 100 K from experimental values.
The high N\'{e}el temperature around 1000 K indicates the thermal stability of the magnetic structure, possibly enabling magnetic devices at room temperature.

Finally, we compute the optical and magneto-optical properties of MnPt via the dielectric function and the reflectivity spectrum including intra- and interband contributions, to provide insight into the underlying physics and the possibility of magneto-optical detection of collective spin motion.
From this, we predict the generation of polar magneto-optical effects of antiferromagnetic MnPt when applying an external magnetic field.
Our simulations show a polar MOKE signal on the order of $\mu$rad for an external field of 1\,T.
We find this to be in the linear response regime and our data can provide guidance for maximizing the polar MOKE signal in experiments with a few hundreds nrad of resolution through linear interpolation.

\begin{acknowledgements}
This work was undertaken as part of the Illinois Materials Research Science and Engineering Center, supported by the National Science Foundation MRSEC program under NSF Award No.\ DMR-1720633.
This work made use of the Illinois Campus Cluster, a computing resource that is operated by the Illinois Campus Cluster Program (ICCP) in conjunction with the National Center for Supercomputing Applications (NCSA) and which is supported by funds from the University of Illinois at Urbana-Champaign.
This research is part of the Blue Waters sustained-petascale computing project, which is supported by the National Science Foundation (awards OCI-0725070 and ACI-1238993) and the state of Illinois.
Blue Waters is a joint effort of the University of Illinois at Urbana-Champaign and its National Center for Supercomputing Applications.
\end{acknowledgements}

\bibliographystyle{apsrev}
\bibliography{main.bib}

\end{document}